\documentclass[conference]{IEEEtran}
\usepackage{cite}
\usepackage{amsmath,amssymb,amsfonts}
\usepackage{algorithmic,enumitem}
\usepackage{graphicx}
\usepackage{pdfpages}
\usepackage{textcomp}
\usepackage{xcolor}
\usepackage{multicol}
\usepackage{lipsum}
\usepackage{mwe}
\usepackage{subfigure}
\usepackage{subcaption, ragged2e}
\usepackage{diagbox}
\usepackage{float}
\usepackage{dblfloatfix}
\usepackage{multicol}
\usepackage{multirow}
\usepackage{graphicx} 
\usepackage{caption} 
\usepackage{subcaption} 
\usepackage{booktabs} 
\usepackage{caption}  
\usepackage[top=0.75in, bottom=1.05in, left=0.625in, right=0.625in]{geometry}
\ifCLASSINFOpdf
\else
\fi

\begin{document}

%

\title{Robust Lightweight IoT Device Fingerprinting Under Network Traffic Anomalies

{\normalfont\footnotesize
\parbox{0.95\textwidth}{%
\centering
\text{Author pre-print copy. The manuscript has been accepted for publication at MILCOM 2026.}


\copyright~2026 IEEE. Personal use of this material is permitted.
Permission from IEEE must be obtained for all other uses,
in any current or future media, including reprinting/republishing
this material for advertising or promotional purposes,
creating new collective works, for resale or redistribution
to servers or lists, or reuse of any copyrighted component
of this work in other works.
}%
}%
}



\author{\IEEEauthorblockN{Roya Taheri, Shamik Sengupta}
\IEEEauthorblockA{Department of Computer Science and Engineering, University of Nevada, Reno, USA\\
1664 N. Virginia Street, Reno, NV 89557\\
rtaheri@unr.edu, ssengupta@unr.edu}
\and
\IEEEauthorblockN{}
\IEEEauthorblockA{ \\
}
 }






\vspace{-4mm}

\maketitle

\vspace{-4mm}
\begin{abstract}
Accurate IoT device identification is essential for network security, yet many existing approaches rely on payload inspection, feature engineering, or long observation windows that limit scalability and often assume stable network conditions despite the inherent variability of real-world IoT traffic. We propose a lightweight IoT device identification approach based on packet headers and locality-sensitive hashing (LSH), generating compact fingerprint representations at packet- and window-level granularities. To evaluate robustness, we consider packet-level traffic anomalies including drop, perturbation, reordering, duplication, and packet injection. Experimental results show that the proposed approach achieves up to 99\% accuracy while maintaining robustness under realistic traffic distortions using short bounded observation windows. Anomalies affecting global traffic structure produce the largest performance degradation, whereas window-level representations improve robustness by reducing sensitivity to localized distortions. The low computational overhead of the proposed approach enables scalable IoT device identification in resource-constrained environments.
\end{abstract}


%
\IEEEpeerreviewmaketitle

\section{Introduction}

The rapid growth of Internet of Things (IoT) devices has significantly expanded the attack surface of modern networks, creating new challenges for device identification and security enforcement. IoT devices are often heterogeneous, resource-constrained, and poorly managed, making traditional authentication and access control mechanisms difficult to deploy effectively. As a result, accurate device identification has become an important component for enabling network security functions such as intrusion detection, anomaly detection, and policy enforcement \cite{atzori2010internet, sicari2015security, lyon2026benchmarking}.

A large body of work has explored IoT device fingerprinting using network traffic characteristics. Early systems such as IoT Sentinel demonstrated that device types can be identified from features extracted from network flows, enabling automated policy enforcement \cite{miettinen2017iot}. Subsequent approaches applied machine learning techniques using statistical features, temporal patterns, and protocol-level behaviors to improve classification accuracy \cite{meidan2017detection, acar2020peek}. Although these methods achieve strong performance under controlled conditions, many rely on flow-level aggregation or detailed feature engineering, resulting in increased computational overhead and limiting deployment in resource-constrained environment.

To address these limitations, recent work has explored lightweight fingerprinting techniques based on compact traffic representations, including locality-sensitive hashing. Similarity-preserving hashing methods such as Nilsimsa enable efficient comparison of traffic patterns without requiring explicit feature extraction \cite{damiani2004open}. These approaches are particularly well suited for real-time and edge-based deployments due to their low computational and memory requirements.

While previous works address those limitations, real-world IoT traffic is highly dynamic and frequently affected by packet-level network variability, including packet loss, retransmissions, reordering, transient congestion, wireless instability, and protocol-level traffic anomalies \cite{sicari2015security,kotak2022adversarial}. 
These effects can alter
traffic structure and reduce the stability of fingerprint representations, leading to degraded identification robustness. This challenge is particularly important in large-scale and resource-constrained edge deployments, where lightweight identification systems must maintain robustness under realistic network conditions while operating on limited observation windows.

In this paper, we investigate the robustness of header-based IoT device fingerprinting under realistic network distortions. We propose a lightweight fingerprint approach using packet headers and Nilsimsa hashing, and evaluate performance across both packet-level and window-level representations.

The main contributions of this work are as follows:
\begin{itemize}
\item A lightweight and scalable IoT device identification approach based on packet headers, locality-sensitive hashing, and lightweight machine learning classifiers, avoiding payload inspection and explicit feature engineering.
\item A systematic robustness evaluation under realistic packet-level network anomalies, including drop, perturbation, reordering, duplication, and packet injection, together with an analysis of representation stability across packet-level and window-level granularities.
\item An overhead analysis demonstrating that compact hash representations enable accurate, resource-efficient IoT device identification in resource-constrained environments.
\end{itemize}

\begin{figure}[h]
    \centering
    \includegraphics[width=0.95\columnwidth]{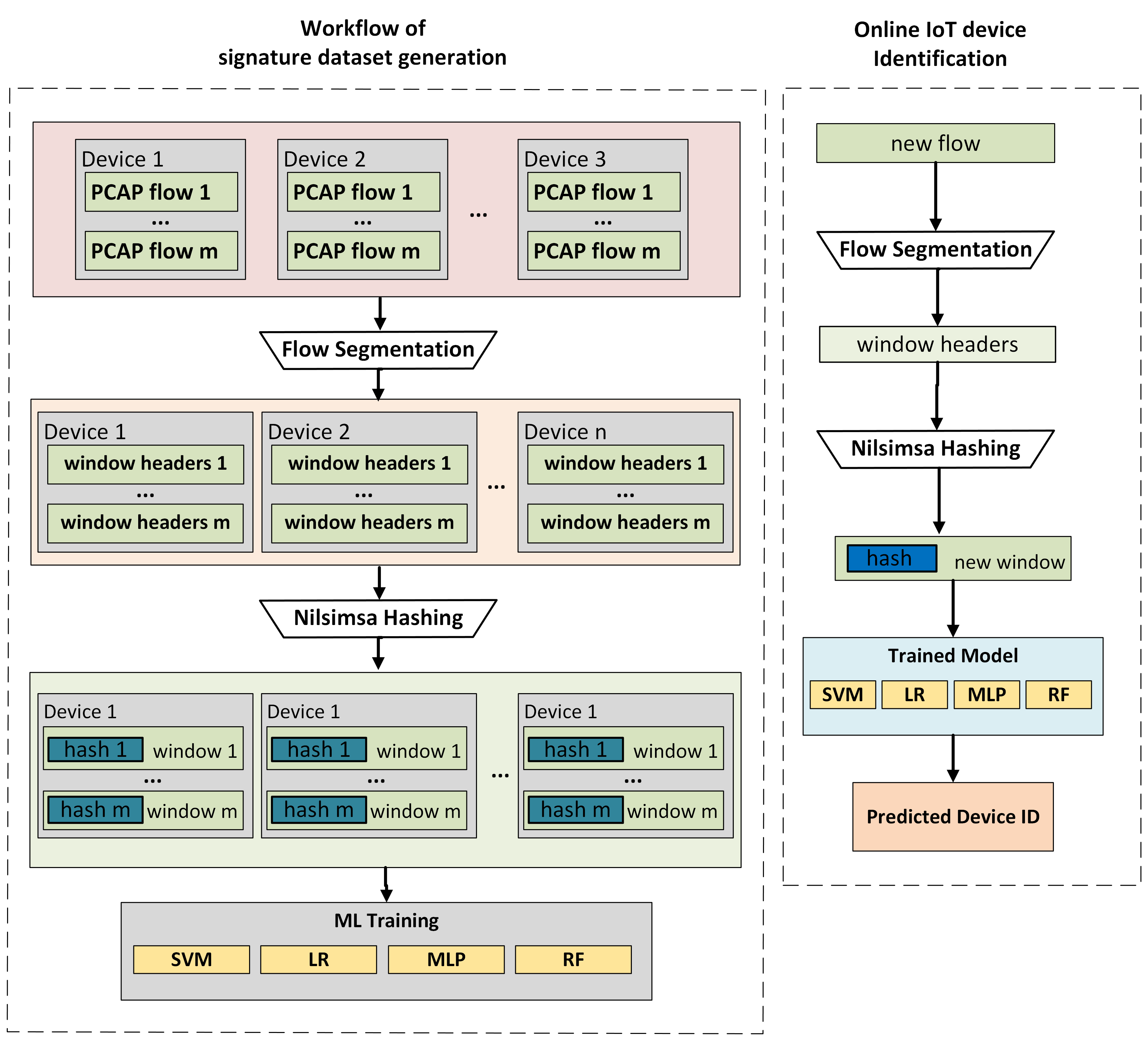}
    \caption{
Overview of the proposed device fingerprinting 
}
    \vspace{-3mm}
    \label{fig:arch}
\end{figure}

\section{Related Work}

IoT device identification has been extensively studied using network traffic analysis and machine learning techniques. Early approaches rely on handcrafted feature extraction combined with classical classifiers. For instance, works such as \cite{aksoy2019automated, meidan2017profiliot, chowdhury2020network} utilize statistical features derived from packet headers and flow-level characteristics to distinguish devices. Similarly, lightweight behavioral-based approaches have been proposed to improve efficiency in resource-constrained environments \cite{pandey2025lightweight, salman2022machine}. While these methods achieve reasonable accuracy, they require careful feature engineering and often struggle to generalize across diverse network conditions.

To reduce reliance on feature design, deep learning-based approaches have been introduced for IoT device identification. Models such as convolutional neural networks (CNNs) and deep neural networks automatically learn discriminative representations from raw traffic data \cite{kotak2019iot, jamali2025iot, qing2020device}. These approaches have demonstrated high accuracy and strong generalization capabilities. However, they typically require large datasets, high computational resources, and long observation windows, limiting their applicability in real-time and resource-constrained IoT environments.

To address these limitations, recent work has explored lightweight and efficient frameworks for IoT identification and security. Several studies focus on reducing computational overhead through simplified models or optimized feature extraction \cite{punithavathi2019lightweight, pandey2025lightweight, rouf2024flexinet}. Although these approaches improve efficiency, they still depend on training and may be sensitive to variations in traffic patterns.
More recently, locality-sensitive hashing (LSH) has emerged as a promising alternative for lightweight IoT fingerprinting. LSH-based methods generate similarity-preserving representations of network traffic, enabling efficient comparison without extensive feature engineering or training. Several works have applied LSH for anomaly detection and traffic identification \cite{charyyev2024identifying, hernandez2024machine, charyyev2020detecting}. The LSIF framework further extends this idea by using Nilsimsa hashing to generate fingerprints for IoT device identification \cite{charyyev2020locality}. Additional works, such as \cite{charyyev2020iot, ming2024iot}, apply LSH to traffic flows or periodic patterns to improve identification accuracy.
Building on this direction, hybrid and enhanced hashing-based approaches have been proposed. SmartRecon combines Nilsimsa hashing with neural networks to enable identification from limited traffic samples \cite{thom2022smart}. Other works improve hashing-based fingerprinting by incorporating behavioral or statistical features \cite{hernandez2024enhancing}. More recent studies, including FlexHash and header-based LSH fingerprinting, introduce tunable hashing mechanisms and multi-layer header representations to improve discrimination and scalability \cite{thom2024flexhash, taheri2026lightweight}. These approaches show that hashing-based methods can achieve high accuracy while maintaining low computational overhead.

Existing IoT fingerprinting approaches primarily rely on statistical traffic features, protocol behavior, or deep learning models trained on packet payloads and long traffic flows. IoT Sentinel \cite{miettinen2017iot} utilizes statistical characteristics extracted from device initialization traffic, while SmartRecon \cite{thom2022smart} combines neural representations with locality-sensitive hashing for scalable device identification. More recently, LSIF \cite{charyyev2020iot} demonstrated that locality-sensitive hashing can effectively identify IoT devices without explicit feature engineering by generating similarity-preserving fingerprints from network traffic.

Despite these advances, most prior work evaluates identification performance under clean traffic conditions. Robustness against realistic packet-level traffic anomalies remains largely unexplored. In contrast, the proposed approach combines lightweight header-based hashing with machine learning classifiers and systematically evaluates robustness and representation stability under packet-level network anomalies.

\section{Methodology}

This section presents the proposed approach, including the overall pipeline, header-based fingerprint generation, and the threat model used for robustness evaluation.
\subsection{Overview of the proposed approach}
Fig.~\ref{fig:arch} illustrates the overall pipeline of the proposed IoT device fingerprinting approach. Raw network traffic in the form of PCAP traces is first processed to extract packet headers across protocol layers, which are then encoded as byte sequences and transformed into fixed-length, similarity-preserving fingerprints using Nilsimsa hashing. To maintain a lightweight and bounded-observation design, fingerprint representations are generated from limited packets of each traffic trace rather than the complete capture. Fingerprint generation is performed at multiple levels of granularity, including packet-level and window-level representations, enabling both fine-grained and aggregated characterization of device behavior. The resulting fingerprint vectors are used as inputs to machine learning classifiers to identify devices from packet-level and window-level traffic representations.

\subsection{Fingerprint Generation}

To maintain a lightweight design, the proposed approach relies exclusively on packet header information. For each packet, protocol headers are extracted and encoded as a byte sequence. By excluding payload data, the approach remains applicable to encrypted traffic and avoids variability introduced by payload content.
To capture device behavior at different temporal scales, fingerprint generation is performed at two levels of granularity. At the packet level, each packet is represented independently, enabling fine-grained characterization. At the window level, consecutive packets are grouped into fixed-size non-overlapping windows to capture short-term communication patterns while maintaining bounded observations. This allows the representation to capture both local packet characteristics and temporal traffic structure.
The resulting byte sequences are transformed into fixed-length fingerprints using locality-sensitive hashing (LSH). Specifically, we employ Nilsimsa hashing \cite{damiani2004open}, which generates similarity-preserving digests for input data. Unlike cryptographic hash functions that produce entirely different outputs for small input changes, Nilsimsa produces similar fingerprints for structurally similar inputs, making it well suited for IoT traffic fingerprinting under network variability \cite{charyyev2020iot, charyyev2020locality}.

\begin{figure}[h]
    \centering
    \vspace{-4mm}
    \includegraphics[width=0.60\columnwidth]{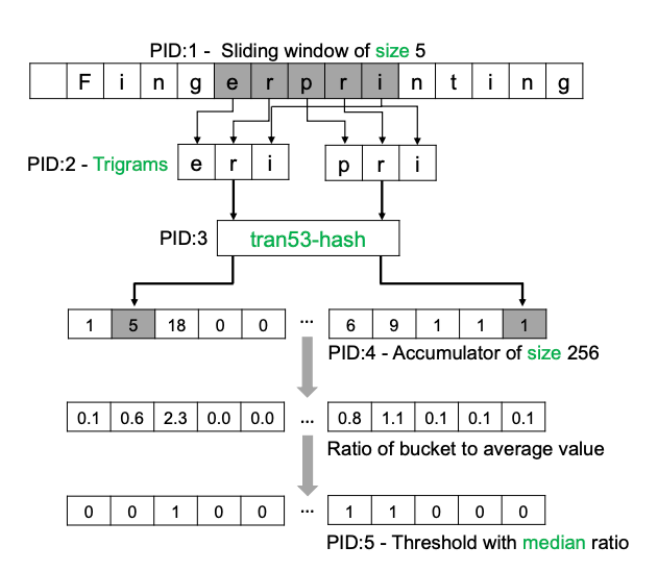}
    \caption{Nilsimsa functionality \cite{charyyev2020locality}}
    \label{fig:nilsimsa}
    \vspace{-4mm}
\end{figure}

As illustrated in Fig.~\ref{fig:nilsimsa}, Nilsimsa processes the input byte sequence using a sliding window to generate overlapping three-byte patterns (trigrams). Each trigram is mapped to one of 256 accumulator positions, where the corresponding counter is incremented. After processing the entire sequence, the accumulator summarizes the distribution of local byte patterns within the traffic representation. The accumulator values are then thresholded to generate a fixed-length 32-byte digest that serves as the final fingerprint. Since similar byte sequences produce similar accumulator distributions, the resulting fingerprints preserve structural similarity while remaining compact and efficient to compare.

\subsection{Threat Model}

We consider a network-level setting in which a monitoring system attempts to identify IoT devices from observed traffic patterns under realistic network variability and packet-level traffic anomalies. We assume that anomalies occur at inference time without modifying training data, compromising end devices, or accessing payload contents. The evaluation focuses on analyzing how packet-level traffic distortions affect fingerprint representation stability and device identification reliability.
Let a traffic trace be represented as an ordered packet sequence $P = \{p_1, p_2, \dots, p_n\}$. The adversary applies a 
function $\mathcal{A}(\cdot)$ to generate a modified sequence $\tilde{P} = \mathcal{A}(P)$ through packet-level traffic anomalies intended to alter the resulting fingerprint representation. Anomalies are applied over localized burst regions $B \subseteq \{1, \dots, n\}$, where $|B| = \lfloor \alpha n \rfloor$ and $\alpha \in [0,1]$ denotes the anomaly ratio. The objective is to evaluate how packet-level traffic anomalies affect representation stability and classification accuracy while preserving protocol-valid traffic behavior.
To evaluate robustness under realistic network distortions, we consider packet-level anomalies including packet drop, header perturbation, packet reordering, packet injection, and packet duplication. Anomalies are applied as temporally localized bursts rather than uniformly across the sequence to model realistic transient network behavior observed in operational IoT networks \cite{sicari2015security,kotak2022adversarial}. Packet \textit{drop} simulates congestion or wireless instability, \textit{perturbation} modifies selected protocol header fields while preserving protocol validity, packet \textit{reorder} models temporal disorder caused by routing dynamics or jitter, packet \textit{injection} introduces protocol-aware spoofed communication behavior, and \textit{duplication} models retransmissions and altered packet frequency distributions.

All anomalies are applied prior to fingerprint generation to emulate realistic deployment conditions where traffic anomalies occur before feature extraction and hashing. Representation stability is quantified using the normalized Hamming distance between clean and manipulated fingerprint representations \cite{hamming1950error}:

\[
D_H(h,\tilde{h}) =
\frac{1}{L}
\sum_{i=1}^{L}
\mathbf{1}[h_i \neq \tilde{h}_i],
\]

where $L$ is the fingerprint lengthcorresponding to the size of the accumulator array. Table~\ref{tab:hash_shift} reports representative hash-space deviations under packet-level anomalies. Reordering produces the smallest representation distortion, whereas injection and combined anomalies generate the largest hash-space deviations. Larger aggregation windows consistently reduce fingerprint distortion across all evaluated anomalies. The limited impact of packet reordering can be attributed to the preservation of packet contents and local trigram statistics despite temporal displacement. Since Nilsimsa accumulates overlapping byte-level trigram distributions rather than strict sequential semantics, localized packet reordering affects only a subset of accumulator positions.



\begin{table}[t]
\centering
\caption{Average normalized Hamming distance between clean and anomalous fingerprints across window sizes.}
\label{tab:hash_shift}
\scalebox{1.05}{
\begin{tabular}{lccccc}
\toprule
\textbf{Anomaly} & \textbf{Ratio} & \textbf{W=1} & \textbf{W=5} & \textbf{W=10} & \textbf{W=20} \\
\midrule\midrule
Reordering & 0.5 & 0.220 & 0.171 & 0.144 & 0.120 \\
Perturbation & 0.5 & 0.230 & 0.185 & 0.158 & 0.132 \\
Injection & 0.5 & 0.231 & 0.188 & 0.162 & 0.141 \\
\bottomrule
\end{tabular}}
\end{table}

\begin{table}[t]
\centering
\caption{Average Classification Accuracy (Mean $\pm$ Std) under Different Window Sizes (No Anomaly)}
\label{tab:accuracy_std}
\scalebox{0.90}{%
\begin{tabular}{ lcccc}
\toprule
\textbf{Model} & \textbf{W=1} & \textbf{W=5} & \textbf{W=10} & \textbf{W=20} \\
\midrule\midrule
LogReg & 0.969 $\pm$ 0.001 & 0.990 $\pm$ 0.002 & 0.994 $\pm$ 0.002 & \textbf{0.999 $\pm$ 0.002} \\
MLP & 0.980 $\pm$ 0.001 & 0.991 $\pm$ 0.001 & 0.994 $\pm$ 0.001 & \textbf{0.998 $\pm$ 0.001} \\
RF & 0.977 $\pm$ 0.001 & 0.983 $\pm$ 0.001 & 0.989 $\pm$ 0.001 & \textbf{0.993 $\pm$ 0.002} \\
SVM & 0.972 $\pm$ 0.002 & 0.986 $\pm$ 0.002 & 0.993 $\pm$ 0.001 & \textbf{0.999 $\pm$ 0.001} \\
\bottomrule
\vspace{-5mm}
\end{tabular}}
\end{table}

\begin{figure*}[t]
\begin{center}

\hfill
\subfigure[W=1]{
    \includegraphics[keepaspectratio=true,angle=0,width=0.23\textwidth]{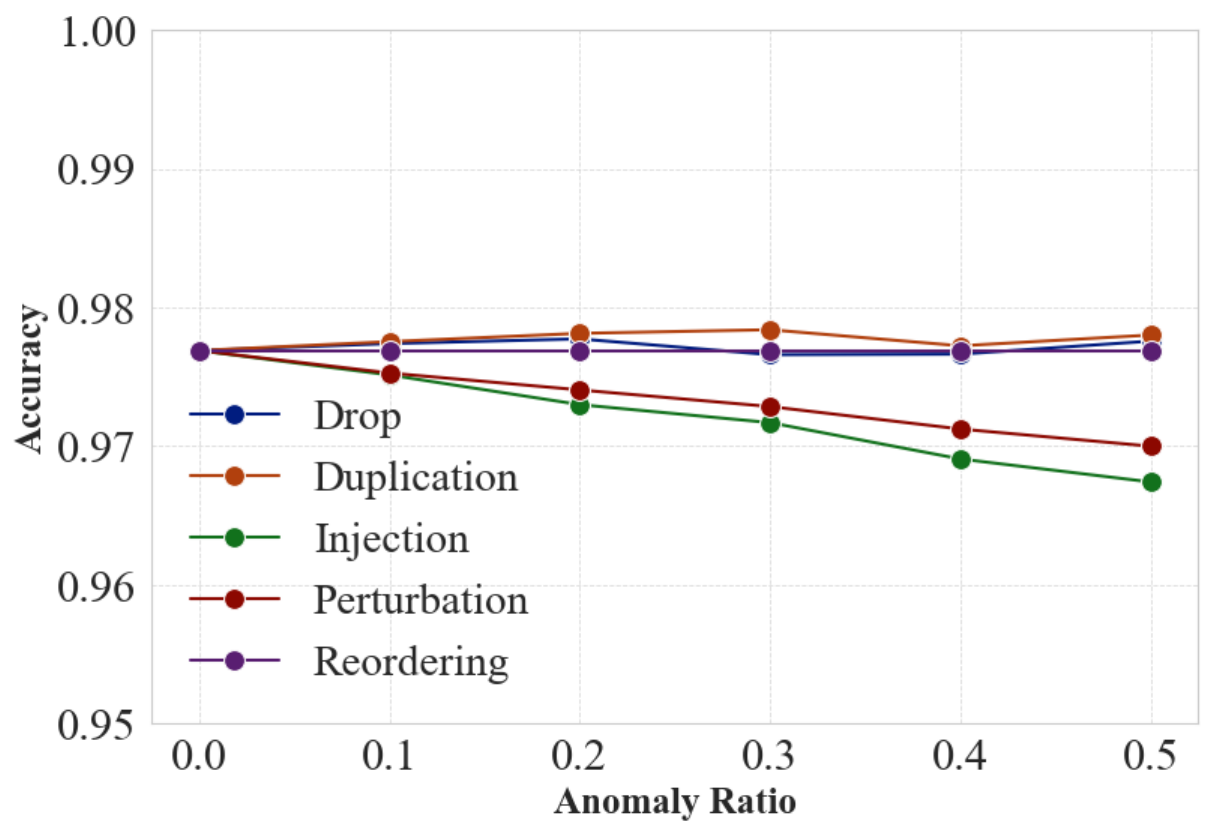}
    \label{fig:fig5}
}
\hfill
\subfigure[W=5]{
    \includegraphics[keepaspectratio=true,angle=0,width=0.23\textwidth]{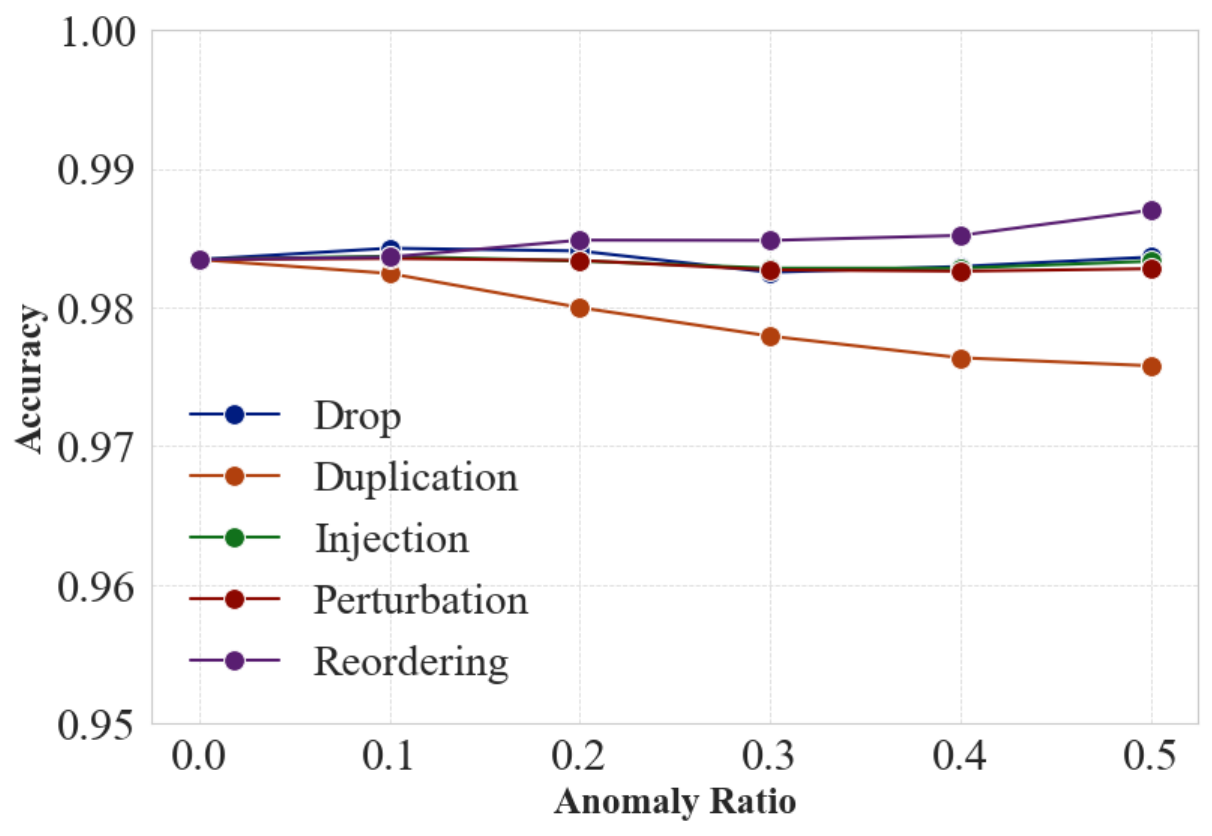}
    \label{fig:fig5}
}
\hfill
\subfigure[W=10]{
    \includegraphics[keepaspectratio=true,angle=0,width=0.23\textwidth]{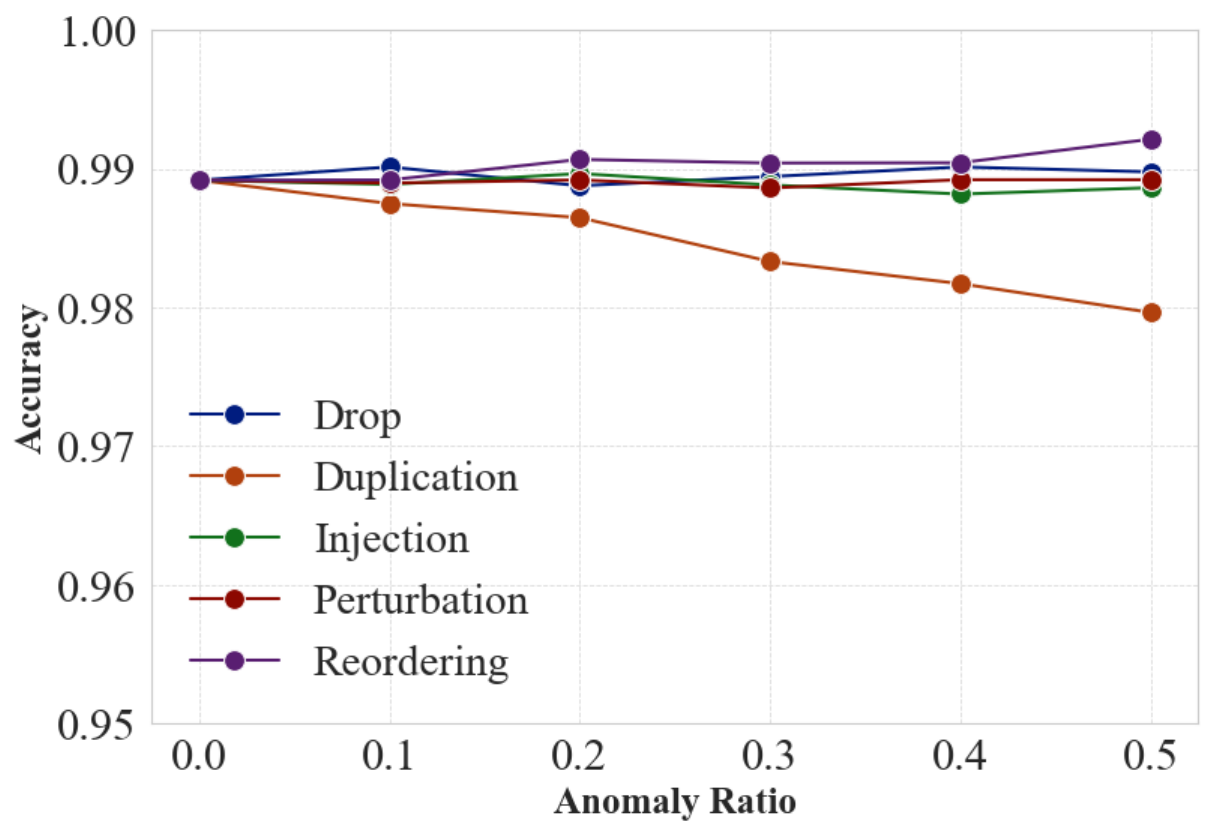}
    \label{fig:fig6}
}
\hfill
\subfigure[W=20]{
    \includegraphics[keepaspectratio=true,angle=0,width=0.23\textwidth]{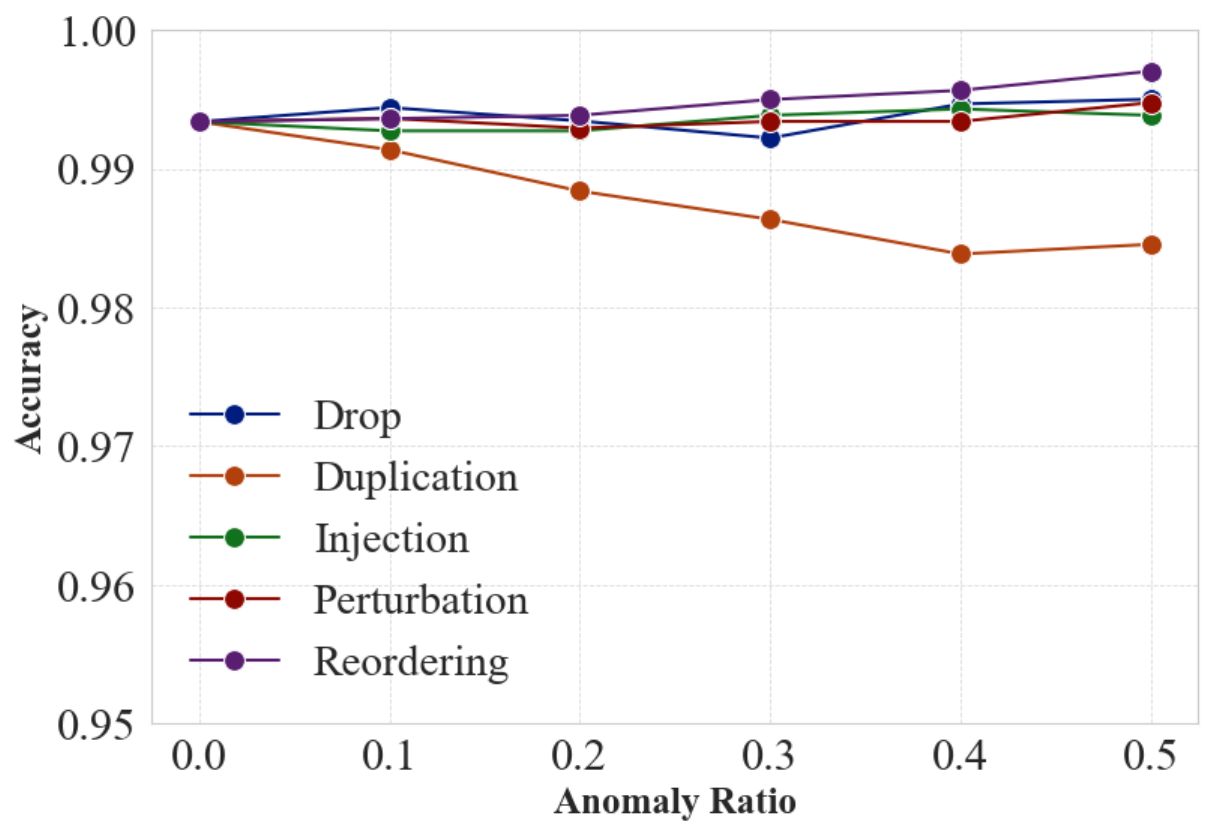}
    \label{fig:fig7}
}

\caption{Classification accuracy versus traffic anomaly ratio for different anomalies using the RF classifier. Injection causes the most significant degradation, while larger window sizes maintain higher accuracy.}
\vspace{-4mm}
\label{fig:attacks}
\end{center}
\end{figure*}

\section{Results and Analysis}

\subsection{Experimental Setup}

We evaluate the proposed fingerprinting approach on the LSIF dataset~\cite{charyyev2020locality}, which includes 22 IoT devices and enables controlled robustness analysis under diverse packet-level transformations. Each device is represented by 20 PCAP traces. From each trace, we extract a fixed lightweight observation window of 200 packets to enable scalable and low-latency device identification under constrained monitoring conditions. The observation size was selected to balance representation stability and processing overhead while avoiding the computational cost associated with long traffic flows. Shorter packet sequences may increase representation variability, whereas substantially larger observation windows introduce additional processing overhead with limited practical benefit for lightweight deployment scenarios.

Traffic is represented at packet-level and window-level granularities. For window-based representations, packets are grouped into fixed-size, non-overlapping windows of size $W \in \{5, 10, 20\}$, and their headers are concatenated prior to hashing.
Fingerprint generation is performed using Nilsimsa hashing, producing fixed-length binary vectors that serve as input features to Random Forest (RF), Support Vector Machine (SVM), Multi-Layer Perceptron (MLP) and Logistic Regression (LR) classifiers. Models are trained on clean traffic and evaluated under both clean and transformed conditions. Traffic anomalies are applied at varying intensity levels, including combined anomaly scenarios.
Performance is evaluated using classification accuracy under a 5-fold cross-validation scheme. The reported mean accuracy corresponds to the average classification accuracy across all five folds, providing a robust estimate of model performance that reduces the impact of variability due to data splitting.
Experiments were conducted on an Ubuntu workstation with an Intel Core i5-7600 CPU @ 3.50 GHz and 32 GB RAM. Fingerprint generation includes packet parsing, header extraction, packet aggregation, and locality-sensitive hashing, while runtime memory usage is measured using process-level monitoring.




\subsection{Robustness Under Different Anomalies}

Under clean conditions, all models achieve high classification accuracy, as shown in Table~\ref{tab:accuracy_std}. Increasing the window size consistently improves performance, with window-level representations ($W \geq 5$) outperforming packet-level hashing ($W=1$). Accuracy increases from approximately 0.97 at $W=1$ to above 0.99 at $W=20$ across models, indicating that aggregation captures more stable device-specific traffic patterns.

As anomaly intensity increases, performance degrades at different rates depending on the anomaly type and representation granularity. Fig.~\ref{fig:attacks} shows classification accuracy as a function of anomaly ratio for $W=1$, $W=5$, $W=10$, and $W=20$. 

At low attack ratios ($\leq 0.1$), all representations maintain high accuracy, indicating robustness to minor traffic distortions. As the anomaly ratio increases, injection and perturbation have the greatest impact on packet-level representations, while duplication becomes the dominant source of degradation for larger window sizes. In contrast, reordering consistently has a negligible effect across all configurations. Overall, larger window sizes maintain higher accuracy and exhibit greater resilience to increasing anomaly intensity, suggesting that packet aggregation helps preserve stable device-specific traffic patterns under localized modifications.

\begin{figure}[h]
    \centering
    \includegraphics[width=0.80\columnwidth]{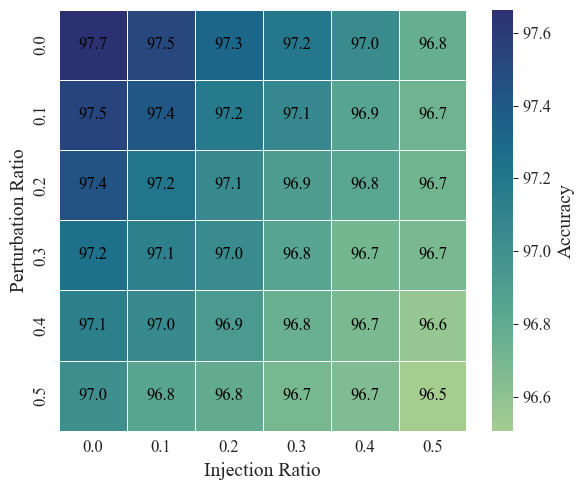}
    \caption{Classification accuracy under combined perturbation and injection ratios using RF classifier.}
    \label{fig:heatmap}
\end{figure}

Figure~\ref{fig:heatmap} presents classification accuracy under combined perturbation and injection anomalies at packet-level granularity. Accuracy gradually decreases as both anomaly intensities increase but remains above 96\% across all evaluated combinations. This suggests that simultaneous localized modifications do not accumulate sufficiently to eliminate the device-specific similarity captured by the fingerprint.


Most devices achieve near-perfect precision across all evaluated window configurations. For clarity and readability, Fig.~5 presents representative devices exhibiting imperfect precision or the largest variation across window sizes. While most devices maintain high precision under packet-level anomalies, several devices exhibit more noticeable degradation at the packet level. Standard deviation across folds remained low for all evaluated devices, indicating stable per-device classification behavior across different train-test partitions. Larger aggregation windows generally improve robustness consistency by reducing sensitivity to localized traffic distortions and transient packet modifications, particularly for devices with more variable traffic patterns. Precision, recall, and F1-score exhibit trends consistent with classification accuracy across different window sizes and traffic anomalies.

Overall, classification degradation follows changes in fingerprint stability: anomalies producing larger normalized Hamming distances, particularly injection-based modifications, cause greater accuracy loss, whereas reordering produces smaller hash-space deviations and limited performance impact. This relationship indicates that robustness is governed primarily by the stability of the similarity-preserving representation rather than classifier complexity alone.

\begin{figure}[h]
    \centering
    \includegraphics[width=0.90\columnwidth]{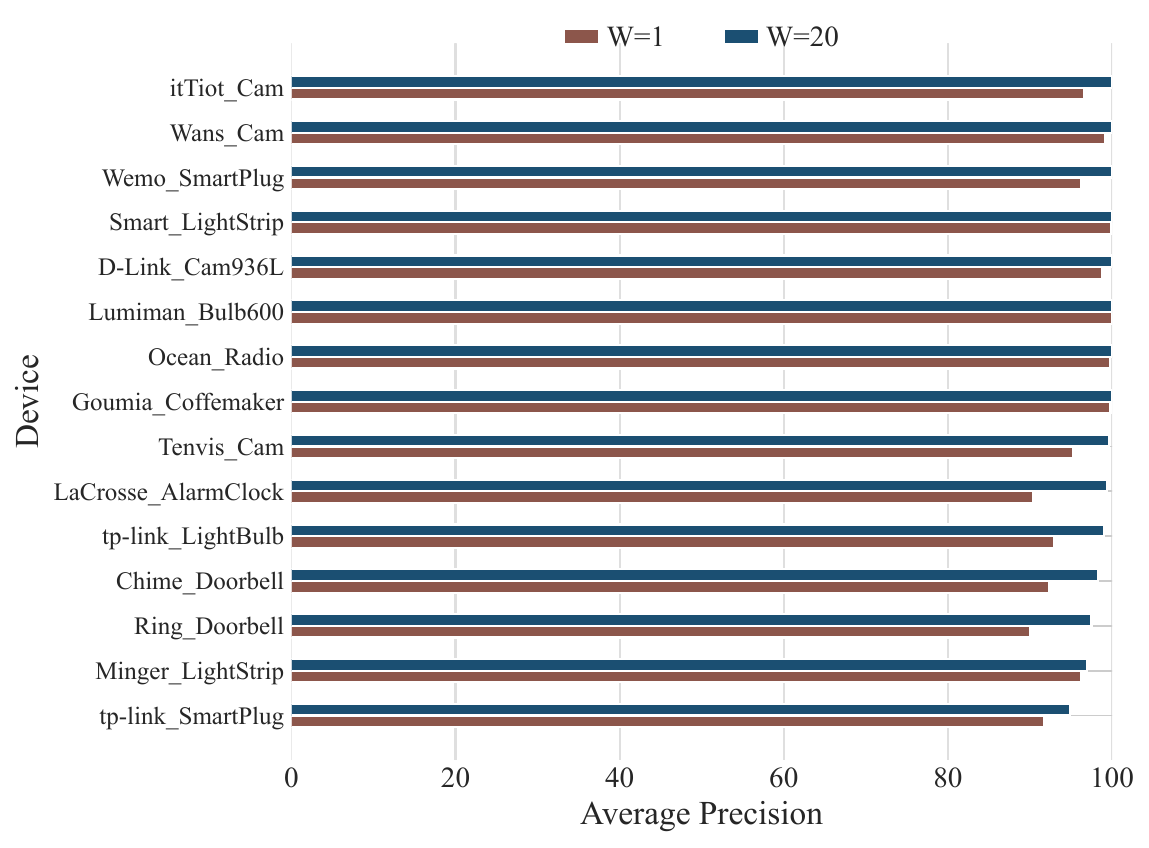}
    \caption{Per-device precision for selected devices using the RF classifier under injection-based traffic anomalies averaged across all injection ratios for packet-level and window-level fingerprint representations.}
    \label{fig:precision}
\end{figure}

\subsection{Overhead Analysis}

We evaluate fingerprint generation overhead, inference latency, and runtime memory usage, as summarized in Table~\ref{tab:hash_overhead} and Table~\ref{tab:overhead}. 
Results show that fingerprint generation remains lightweight across all evaluated window sizes. Hashing latency increases with larger aggregation windows due to additional packet concatenation and trigram processing, rising from 0.91\,ms at $W=1$ to 14.34\,ms at $W=20$. However, memory overhead remains nearly constant at approximately 1\,KB across all configurations, reflecting the fixed-size accumulator structure of Nilsimsa hashing. Inference overhead also remains low across all evaluated models. Linear classifiers such as LR and SVM achieve particularly low inference latency, while RF incurs moderately higher overhead due to ensemble-based decision traversal. Runtime memory usage remains relatively stable across configurations, with RF exhibiting the highest overhead among the evaluated models.

Overall, the proposed approach maintains low computational overhead while achieving robust identification performance under packet-level anomalies, making it suitable for scalable edge-based IoT device identification in resource-constrained environments.

\begin{table}[t]
\centering
\caption{Fingerprinting overhead across window sizes.}
\label{tab:hash_overhead}
\scalebox{1.0}{
\begin{tabular}{lcc}
\toprule
\textbf{Window} &
\textbf{Time (ms)} &
\textbf{Memory (B)} \\
\midrule\midrule
W=1  & $0.91 \pm 0.54$ & $913 \pm 287$ \\
W=5  & $3.76 \pm 2.31$ & $938 \pm 246$ \\
W=10 & $7.34 \pm 4.34$ & $953 \pm 225$ \\
W=20 & $14.34 \pm 8.20$ & $934 \pm 154$ \\
\bottomrule
\end{tabular}}
\vspace{-2mm}
\end{table}

\begin{table}[H]
\centering
\caption{Inference overhead per fingerprint representation.}
\label{tab:overhead}
\scriptsize
\begin{tabular}{lcc}
\toprule
\textbf{Model} &
\textbf{Latency (ms)} &
\textbf{Runtime Memory (MB)} \\
\midrule\midrule

LogReg & 0.06 & 349 \\
MLP    & 0.10 & 353 \\
RF     & 4.01 & 393 \\
SVM    & 0.06 & 353 \\

\bottomrule
\end{tabular}
\end{table}

\subsection{Comparison with Prior Approaches}

Unlike prior approaches, the proposed method combines lightweight header-based locality-sensitive hashing with bounded packet observations and lightweight machine learning classifiers. Table~V compares representative IoT fingerprinting approaches in terms of feature representation, robustness evaluation, observation requirements, and reported accuracy. While existing approaches either rely on initialization traffic, long traffic flows, or do not explicitly evaluate robustness under network variability, the proposed approach systematically evaluates performance under realistic packet-level anomalies. Despite relying only on packet headers and limited packet observations, it achieves the highest reported accuracy among the compared approaches while maintaining low computational overhead, making it suitable for deployment in resource-constrained environments.

\begin{table}[H]
\centering
\caption{Comparison with existing IoT fingerprinting approaches.}
\label{tab:comparison}
\scalebox{1.0}{
\begin{tabular}{lcccc}
\toprule
Method & Features & Eval. & Obs. & Acc. \\
\midrule\midrule
IoT Sentinel \cite{miettinen2017iot} & Statistical & No & Setup & 95\% \\
LSIF \cite{charyyev2020iot} & LSH+Similarity & No & Flows & 90--93\% \\
SmartRecon \cite{thom2022smart} & LSH+Neural & No & Limited & 94\% \\
Proposed & Header-LSH+ML & Yes & Limited & 99\% \\
\bottomrule
\end{tabular}}
\end{table}

Although the evaluation focuses on the LSIF dataset, the proposed approach is designed to remain independent of device-specific payload content and protocol semantics. Evaluating the proposed approach on additional IoT datasets is an important direction for future work.


\section{Conclusion}

We presented a lightweight IoT device fingerprinting approach based on packet headers and LSH. The proposed method achieves high identification accuracy while avoiding payload inspection, explicit feature engineering, and large observation windows. Experimental results show that transformations affecting global traffic structure produce the largest performance degradation, whereas window-level representations improve robustness by reducing sensitivity to localized distortions. In addition, the proposed approach maintains low computational and memory overhead through compact fixed-length hash representations and bounded observation windows, making it suitable for scalable edge-based IoT device identification under realistic packet-level network variability.
\section*{Acknowledgments}
This study was supported by the NSF under grant 2346755.
\bibliographystyle{IEEEtran}
\bibliography{main}

\end{document}